\documentclass[aip,jcp,preprint,unsortedaddress]{revtex4-2}

\usepackage{graphicx}
\usepackage{dcolumn}
\usepackage{bm}
\usepackage{amsmath,amssymb}
\usepackage{multirow}
\usepackage[usenames,dvipsnames]{xcolor}
\usepackage{overpic}
\usepackage{pict2e}
\usepackage{mathtools}
\usepackage{gensymb}
\usepackage{physics}
\usepackage{siunitx}
\usepackage{hyperref}
\usepackage{multirow}

\newcommand{\scrm}[1]{\mbox{\scriptsize #1}}

\definecolor{ggreen}{rgb}{0.0, 0.5, 0.0}

\newcommand{\omu}{Department of Mechanical Engineering, Osaka Metropolitan University, 1-1 Gakuen-tyo, Naka-ku, Sakai, Osaka 599-8531, Japan}
\newcommand{\osaka}{Department of Mechanical Engineering, Osaka University, 2-1 Yamadaoka, Suita, Osaka 565-0871, Japan}
\newcommand{\tokyo}{Water Frontier Science \& Technology Research Center (W-FST), Research Institute for Science \& Technology,
Tokyo University of Science, 1-3 Kagurazaka, Shinjuku-ku, Tokyo 162-8601, Japan}

\begin{document}

\title{Molecular anatomy of the pressure anisotropy in the interface of one and two component fluids: local thermodynamic description of the interfacial tension}

\author{Takeshi Omori}
\email{t.omori@omu.ac.jp}
\affiliation{\omu}
\author{Yasutaka Yamaguchi}
\email{yamaguchi@mech.eng.osaka-u.ac.jp}
\affiliation{\osaka}\affiliation{\tokyo}

\date{\today}

\begin{abstract}
Through the decomposition of the pressure into the kinetic and the intermolecular contributions, we show that the pressure anisotropy in the fluid interface, which is the source of the interfacial tension, comes solely from the latter contribution. The pressure anisotropy due to the intermolecular force between the fluid particles in the same or the different fluid components is approximately proportional to the multiplication of the corresponding fluid density gradients, and from the molecular dynamics simulation of the liquid-vapor and liquid-liquid interfaces, we demonstrate that the density gradient theory (DGT) by van der Waals gives the leading order approximation of the free energy density in inhomogeneous systems, neglecting the Tolman length. 
\end{abstract}

\maketitle

\section{Introduction}
Interface between fluids is ubiquitous, and it is at the heart of many industrial applications to understand its nature and then to tame its dynamical behavior\cite{Scardovelli1999,degennes2002,ghoufi2016}. The dynamical property of the fluid interface is represented by the interfacial tension. From a viewpoint of thermodynamics, by writing the Helmholtz free energy $F$ of a system with a fluid interface of area $A$ as
\begin{equation}\label{eq:basic}
F = -pV 
+ \sum_{\alpha} \mu_{\alpha}N_{\alpha}
+ \gamma A,
\end{equation}
where $p$ and $V$ are the pressure and the volume of the system, $\mu_{\alpha}$ and $N_{\alpha}$ are the chemical potential and the number of particles of a fluid component $\alpha$ in the system, it is understood that the interfacial tension $\gamma$ is the free energy at the interface per area. 
By observing the variation in the system free energy $F$ during the stretching of a planar interface under constant system volume, Bakker\cite{Bakker1928} further related the system pressure to the interfacial tension by
\begin{equation}\label{eq:bakker}
\gamma = \int_{-\infty}^{\infty} [\,p_{\scrm{N}}-p_{\scrm{T}}(n)] \dd{n},
\end{equation} 
where $n$ is the coordinate normal to the interface, and $p_{\scrm{N}}$ and $p_{\scrm{T}}$ are the pressures normal and tangential to the interface, respectively. Here, the pressure is treated as a tensor quantity in contrast to a scalar quantity in Eq.~\eqref{eq:basic}, and $p_{\scrm{T}}$ is position dependent or inhomogeneous, while $p_{\scrm{N}}$ is homogeneous by the mechanical stability between the two phases. 

The thermodynamic description by Eq.~\eqref{eq:basic} is not satisfactory when we wish to predict the distribution of quantities in inhomogeneous systems, especially in dynamic situations\cite{Anderson1998}. Using Eq.~\eqref{eq:bakker} and the fact that the chemical potentials are constant throughout the interface, Eq.~\eqref{eq:basic} can also be written as
\begin{equation}\label{eq:local}
\begin{aligned}[b]
A\int_{-\infty}^{\infty} f(n) \dd{n} 
&= 
A\int_{-\infty}^{\infty}
\left\{
 -p_{\scrm{N}} +
\sum_{\alpha} \mu_{\alpha}\rho_{\alpha}(n) +
[p_{\scrm{N}}-p_{\scrm{T}}(n)]
\right\}\dd{n} \\
&= 
A\int_{-\infty}^{\infty}
\left[
 -p_{\scrm{T}}(n) +
\sum_{\alpha}\mu_{\alpha}\rho_{\alpha}(n)
\right]\dd{n},
\end{aligned}
\end{equation}
where $f(n)$ is the free energy per unit volume and $\rho_{\alpha}(n)$ is the number density of a fluid component $\alpha$, both at $n$. Equation \eqref{eq:local} suggests to write the free energy density as
\begin{equation}\label{eq:f}
f(n) = -p_{\scrm{T}}(n)
+ \sum_{\alpha}\rho_{\alpha}(n)\mu_{\alpha}
\end{equation}
in the form of the local thermodynamics. It is in fact equivalent to the expression suggested earlier by Hill\cite{hill1952}. In a region where there is no interface, the pressure is homogeneous as $p=p_{\scrm{T}}=p_{\scrm{N}}$, and Eq.~\eqref{eq:f} becomes equivalent to the more familar form: 
\begin{equation}
f = -p + \sum_{\alpha}\rho_{\alpha}\mu_{\alpha}.
\end{equation}
The location of the interface in Eq.~\eqref{eq:basic} can be determined to reproduce the total moment of force in the system. It is called the surface of tension (SoT)\cite{ono1960} and given by
\begin{equation}\label{eq:sot}
n_s = 
\frac{1}{\gamma}
\int_{-\infty}^{\infty} n[\,p_{\scrm{N}}-p_{\scrm{T}}(n)] \dd{n}.
\end{equation}
The location of the SoT generally differs from the equimolar surface $n_e$ called the Gibbs dividing surface (GDS). For multi-component systems, there is no unique way to determine the GDS. One way suggested by \citet{Rowlinson1982} is to find $n_e$ that satisfies 
\begin{equation}\label{eq:gds}
\int_{-\infty}^{n_e} \sum_{\alpha}\mu_{\alpha}[\rho_{\alpha}(n)-\rho_{\alpha}(-\infty)] \dd{n}
=
\int_{n_e}^{\infty} \sum_{\alpha}\mu_{\alpha}[\rho_{\alpha}(\infty)-\rho_{\alpha}(n)] \dd{n}.
\end{equation}
The difference between the SoT and GDS, namely $n_e-n_s$, is called the Tolman length\cite{tolman1949}, which is one characteristic length concerning the dynamics of the interface.

Local thermodynamic description of inhomogeneous systems has a long history\cite{Rowlinson1993}. The simplest description was to assume that the free energy density is a function of local thermodynamic variables in the same way as in homogeneous systems\cite{tolman1948,ono1960}, which is distinctively called as point thermodynamics by \citet{Rowlinson1982}. 
In the density gradient theory (DGT) by van der Waals\cite{Rowlinson1979,cahn1958}, the free energy density is expanded in a Taylor series of density gradients and written as, to the most significant order,
\begin{equation}\label{eq:waals}
f = f_0 + 
\frac{\kappa}{2}\left(\dv{\rho}{n}\right)^2,
\end{equation} 
where $f_0$ is the free energy density in the homogeneous system and $\kappa$ is the so-called influence parameter. The interfacial tension is given as the excess free energy by the existence of the interface, and written as\cite{Hansen2013}
\begin{equation}\label{eq:gamma_waals}
\gamma = \int_{-\infty}^{\infty} \kappa \left(\dv{\rho}{n}\right)^2 \dd{n}
\end{equation}
for the DGT. Comparing Eqs.~\eqref{eq:bakker} and \eqref{eq:gamma_waals}, it is now clear that the DGT is equivalent to modeling the pressure anisotropy $p_{\scrm{N}}-p_{\scrm{T}}$ in the interface by the square of the local density gradient:
\begin{equation}\label{eq:p_dgt}
p_{\scrm{N}}-p_{\scrm{T}}(n) = \kappa \left(\dv{\rho}{n}\right)^2.
\end{equation} 
In terms of the local thermodynamics, the DGT offers a mechanism to obtain the pressure in Eq.~\eqref{eq:f}, which is a tensor component in nature. The extension of Eq.~\eqref{eq:gamma_waals} to multi component systems was suggested by \citet{Davis1982} as
\begin{equation}\label{eq:waals_multi}
\gamma 
= 
\sum_{\alpha,\beta}
\int_{-\infty}^{\infty} 
\kappa_{\alpha\beta} \dv{\rho_{\alpha}}{n}\dv{\rho_{\beta}}{n} \dd{n},
\end{equation}
where the summation is over all the components. The pressure anisotropy in multi component systems can also be expressed with the density gradients, which we will show in \ref{sec:results} [Eq.~(\ref{eq:dgt_two})].

In this article, by simple analytical arguments we derive the local thermodynamic description of the pressure anisotropy in an inhomogeneous system for both single  component [Eq.~(\ref{eq:p_dgt})] and multi component systems, which is of the DGT-type. Knowing the description of the pressure anisotropy is equivalent to knowing the description of the inhomogeneous part of the thermodynamic function, i.e.\ the excess free energy density in the interface [Eqs.~(\ref{eq:waals}) to (\ref{eq:p_dgt})]. The key idea is the decomposition of the pressure into the kinetic and intermolecular parts: we show that only the latter part contributes to the pressure anisotropy in an inhomogeneous system and hence to the surface tension. From this argument, it will be made clear why the point thermodynamic description, which does not involve the density gradient, of inhomogeneous systems fails. We also show that the DGT-type local thermodynamic description is an approximation, whose major consequence is the ignorance of the Tolman length of the interface.

\section{Computational Methods}\label{sec:method}
\begin{figure}
\includegraphics[width=\linewidth]{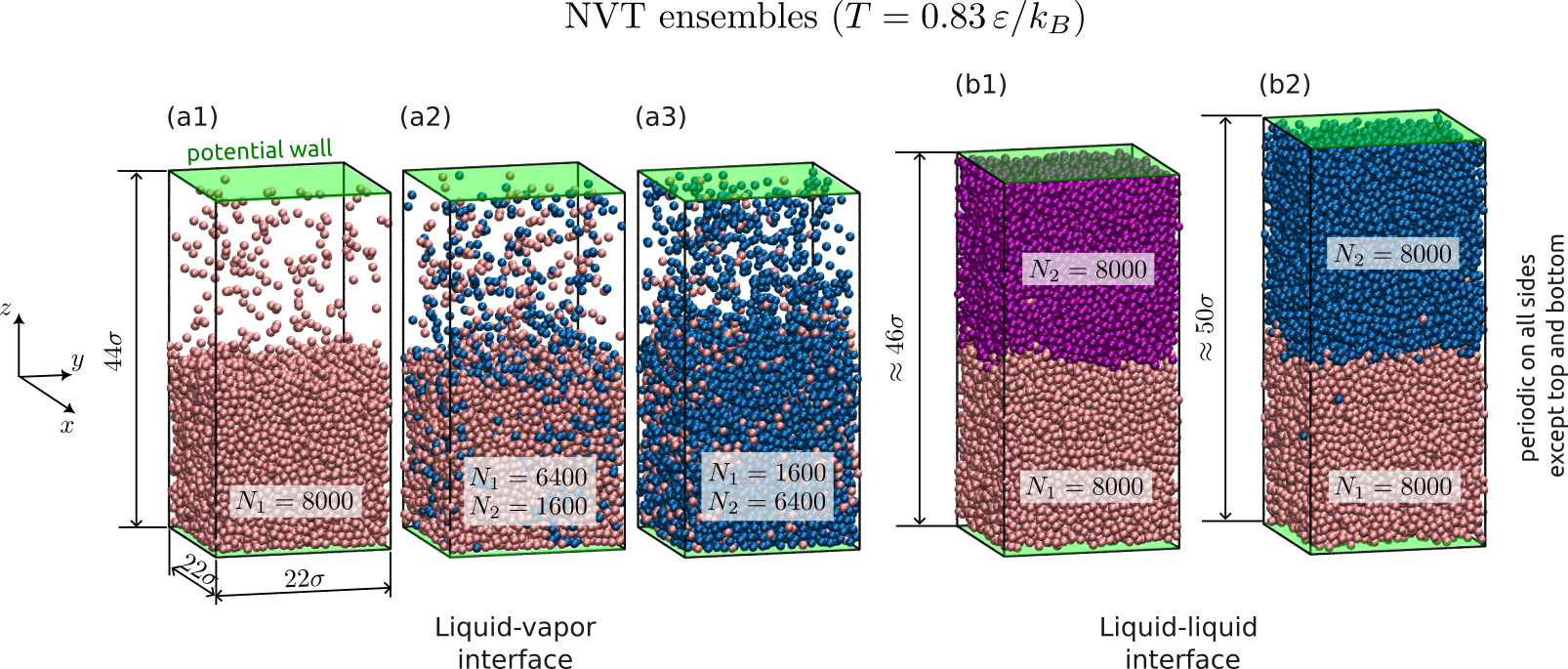}
\caption{Snapshots of equilibrium (a1 to a3) liquid-vapor and (b1 and b2) liquid-liquid systems investigated in the present work. The systems were confined between the potential walls [Eq.~(\ref{eq:w_pot})] on the top and bottom boundaries, and the periodic boundary condition was applied to all side boundaries. The volume of the liquid-liquid system was set so that the bulk pressure of the system was about $9.4\times10^{-2}\varepsilon/\sigma^3$. 
The cross section of the systems was all $(22\times22)\sigma^2$. The origin of $z$-axis was placed at the half height of the systems.}
\label{fig:system}
\end{figure}

To evaluate the pressure distribution in the fluid interfaces, we performed equilibrium MD simulations of liquid slabs confined between two flat walls on the top and bottom and bounded by periodic boundaries on the sides (Fig.~\ref{fig:system}). The interaction between fluid particles was modeled with the 12-6 Lennard-Jones (LJ) potential
\begin{equation}
\Phi_{\scrm{LJ}}(r)=4\varepsilon
\left[
\left(\frac{\sigma}{r}\right)^{12}-
\left(\frac{\sigma}{r}\right)^{6}
\right]
\end{equation}
with $r$ being the inter-particle distance, and $\varepsilon$ and $\sigma$ being the interaction energy and length parameters. We added a quadratic function to $\Phi_{\scrm{LJ}}$ so that both $\Phi_{\scrm{LJ}}(r)$ and $\Phi'_{\scrm{LJ}}(r)$ vanished at the cutoff distance $r_\mathrm{c}=3.5\sigma$\cite{Nishida2014}. The fluids were mixtures of two components: one component of a higher critical temperature (higher $\varepsilon$) was labeled with ``1" and the other of a lower critical temperature (lower $\varepsilon$) with ``2". The length parameter of both components were the same, $\sigma_{11}=\sigma_{22}=\sigma$, and $\sigma_{12}=(\sigma_{11}+\sigma_{22})/2=\sigma$. For the liquid-vapor systems, the parameters for the interaction energy were $\varepsilon_{11}=\varepsilon$ and $\varepsilon_{22}=0.8\varepsilon$ between particles of the same component, and $\varepsilon_{12}=\xi_{12}\sqrt{\varepsilon_{11}\varepsilon_{22}}$ with $\xi_{12}$ being 0.85 between particles of different components. For the liquid-liquid systems, two types of fluid ``2" were investigated: $\varepsilon$ as well as $0.8\varepsilon$ were employed for $\varepsilon_{22}$, and $\xi_{12}$ was set 0.45 (Table~\ref{tab:params}).
The potential between a fluid particle and the confining walls were given by
\begin{equation}\label{eq:w_pot}
\Phi_{\scrm{s}}(h)=
4\pi\rho_s\sigma_{\scrm{fw}}^3\varepsilon_{\scrm{fw}}
\left[
\frac{1}{45}\left(\frac{\sigma_{\scrm{fw}}}{h}\right)^{9}-
\frac{1}{6}\left(\frac{\sigma_{\scrm{fw}}}{h}\right)^{3}
\right]
\end{equation}
with $h$ being the distance of a particle to the wall, and $\rho_s=4/(\sqrt{2}r_0)^3$ being the average number density of solid atoms in a FCC crystal ($r_0=0.91\sigma$), $\sigma_{\scrm{fw}}=1.01\sigma$, and $\varepsilon_{\scrm{fw}}=\varepsilon$ for the bottom wall and $\varepsilon_{\scrm{fw}}=0.155\varepsilon$ for the top wall. 

All the simulations were conducted under constant $NVT$. The system temperature was controlled at $0.83\varepsilon/k_B$ (some calculations were done under temperatures closer to the critical temperature and shown in Appendix \ref{app:temp}) by the Nosé-Hoover chains method\cite{Martyna1992} with the characteristic time scale $\tau$ being $20\Delta t$. Here $\Delta t$ is the time step of the MD simulations and $\Delta t=\num{2.33e-3}\sqrt{m\sigma^2/\varepsilon}$ was employed for all simulations, where $m$ is the mass of the fluid particle, set identical for all fluid components. The equilibrium states of the systems were chosen to be between the triple and critical points of the fluid of both components. The size of the system was large enough to avoid systematic errors\cite{Malfreyt2014,Yamaguchi2019}, with the cross sectional area being $(22\times22)\sigma^2$ and the bulk liquid film being at least $12.5\sigma$ thick excluding the fluid-fluid and fluid-wall interfaces (See also Appendix \ref{app:size}). 

\begin{table}[t]
\caption{Main MD simulation parameters. All the quantities are nondimensionalized with $\sigma$ and $\varepsilon$, whose particular choice does not affect the present results.}\label{tab:params}
\vspace*{1em}
\begin{tabular}{ccc}
\hline\hline
\multirow{2}{*}{Property} & \multicolumn{2}{c}{Value} \\\cline{2-3}
 & Liquid-vapor  & Liquid-liquid \\
\hline
$\sigma_{11}$, $\sigma_{22}$, $\sigma_{12}$ & $\sigma$ & $\sigma$\\
$\varepsilon_{11}$ & $\varepsilon$ & $\varepsilon$\\
$\varepsilon_{22}$ & $0.8\varepsilon$ & $\varepsilon$ [Fig.~\ref{fig:system}(b1)], $0.8\varepsilon$ [Fig.~\ref{fig:system}(b2)] \\
$\varepsilon_{12}$ & $0.85\sqrt{\varepsilon_{11}\varepsilon_{22}}$ & $0.45\sqrt{\varepsilon_{11}\varepsilon_{22}}$ \\
$T$ & $0.83\varepsilon/k_B$ & $0.83\varepsilon/k_B$\\
\hline\hline
\end{tabular}
\end{table}

\subsection{Calculation of the local pressure}
\label{sec:pressure}
The distributions of the pressure tensor and the fluid density in the system were measured by the Method-of-Plane\cite{Todd1995,Kusudo2021}. The spacial resolution of the bin was $0.29\sigma$ for the $z$ direction. The pressure tensor calculated by the Method-of-Plane satisfies the momentum consevation law and the Cauchy definition of pressure, which is an important basis for the discussion of thermodynamics of a system. The pressure tensor $p_{kl}$, which we define as the $l$-th component of the negative stress acting on the plane with its normal pointing in the $k$-th direction, consists of the kinetic and the intermolecular contributions: 
\begin{align}
p_{kl} &= p_{kl}^{\scrm{kin}} + p_{kl}^{\scrm{int}}\\
p_{kl}^{\scrm{kin}} &= 
\frac{1}{S_k \Delta t}
\left<
\sum_{i \in \scrm{fluid}}^{\Delta t, \scrm{crossing }S_k} \frac{m^i v^i_k v^i_l}{|v^i_k|}
\right>\\
p_{kl}^{\scrm{int}} &=
\frac{1}{S_k}
\left<
\sum_{(i,j) \in \scrm{fluid}}^{\scrm{across }S_k} F^{ij}_l \frac{r^{ij}_k}{|r^{ij}_k|}
\right>
\end{align}
with $S$ being the area of the control surface, $\Delta t$ the time step of the MD simulation, $m^i$ and  $v^i$ the mass and the velocity of the fluid particle $i$, $F^{ij}$ and $r^{ij}$ the force and the distance from the fluid particle $i$ to $j$\cite{Kusudo2021}. In the present study, the fluid interface was parallel to the $xy$ plane and therefore $p_{\scrm{N}}=p_{zz}$ and $p_{\scrm{T}}=p_{xx}=p_{yy}$. The distribution of the tangential pressure $p_{\scrm{T}}$ may be slightly different if it is calculated by the contour integral methods (See Appendix \ref{app:temp}), but the following discussions are independent of the choice of the pressure calculation method. The statistics were obtained by averaging the results of 40 statistically independent runs of $\num{4.91e2}\sqrt{m\sigma^2/\varepsilon}$ after the system equilibration, unless otherwise stated. 

\section{Pressure anisotropy in the interface}\label{sec:results}
\subsection{Liquid-vapor interface}
\subsubsection{One component system}
A liquid-vapor interface of one component fluid (See a1 of Fig.~\ref{fig:system}) was first investigated.  
\begin{figure}
\centering
\includegraphics[width=.9\linewidth]{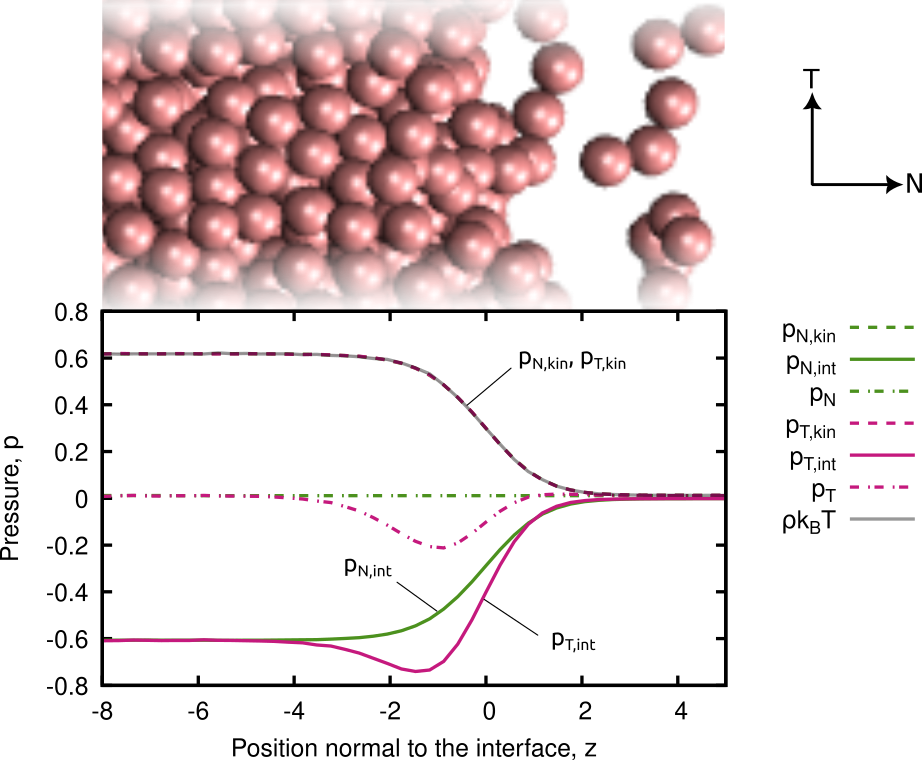}
\caption{Decomposed pressure distributions across the liquid-vapor interface of a one component fluid. The kinetic parts of the normal and tangential pressures overlap. The density profile multiplied by $k_BT$ coincides with the kinetic part of the pressure profile [Eq.~(\ref{eq:virial})].}
\label{fig:one_c}
\end{figure}
Figure \ref{fig:one_c} shows the pressure distribution across the liquid-vapor interface of the one component fluid. Away from the interface, pressure is isotropic ($p_{\scrm{N}}=p_{\scrm{T}}$). As described in \ref{sec:pressure}, pressure can be decomposed into kinetic and intermolecular parts. This decomposition of pressure is similar to the intuitive schematics by \citet{Marchand2011}, but a more rigorous definition is given here. For the normal pressure $p_{\scrm{N}}$, the kinetic part $p_{\scrm{N}}^{\scrm{kin}}$ ($>0$) is everywhere larger in magnitude than the intermolecular part $p_{\scrm{N}}^{\scrm{int}}$ ($<0$) by the saturated vapor pressure, and $p_{\scrm{N}}$ is positive. For the tangential pressure $p_{\scrm{T}}$, there is a region where the intermolecular part exceeds the kinetic part in magnitude and $p_{\scrm{T}}$ gets negative\footnote{From Fig.~\ref{fig:one_c} and Eq.~\eqref{eq:p_dgt2}, $p_{\scrm{T}}$ gets negative when $(\dv*{\rho}{z})^2 \gg \rho_g k_B T$, where the RHS is nearly equal to the saturated vapor pressure with the vapor density $\rho_g$.}. 

The kinetic part of pressure is everywhere isotropic ($p_{\scrm{N}}^{\scrm{kin}}=p_{\scrm{T}}^{\scrm{kin}}$), because the following relationship holds from the classical virial theorem:
\begin{equation}\label{eq:virial}
p^{\scrm{kin}}=\rho k_BT,
\end{equation}
where $\rho$ is the number density of the fluid particle. Then, Eq.~\eqref{eq:bakker} tells that only the intermolecular part contributes to the surface tension:
\begin{equation}
\gamma
=
\int_{-\infty}^{\infty} [\,p_{\scrm{N}}^{\scrm{int}}(z)-p_{\scrm{T}}^{\scrm{int}}(z)] \dd{z},
\end{equation}
with $z$ being the direction of the interface normal $n$. It should be noted that $p_{\scrm{N}}^{\scrm{int}}$ is inhomogeneous although $p_{\scrm{N}}=p_{\scrm{N}}^{\scrm{kin}}+p_{\scrm{N}}^{\scrm{int}}$ is homogeneous, as shown in Fig.~\ref{fig:one_c}. 

\begin{figure}
\includegraphics[width=.9\linewidth]{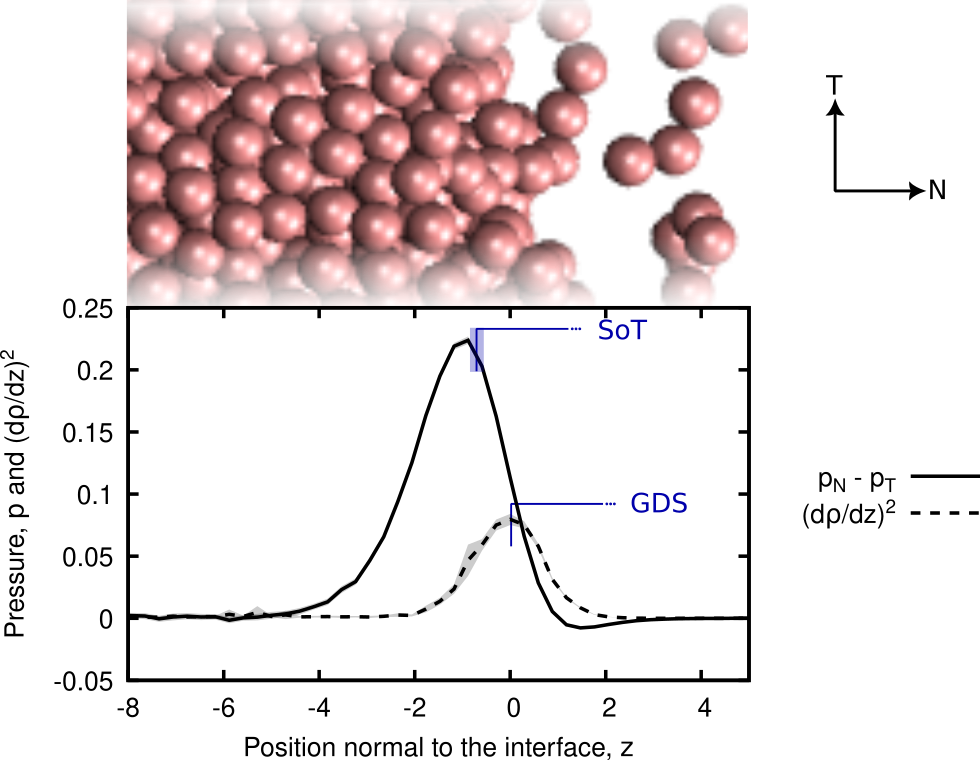}
\caption{Pressure anisotropy and density gradient across the liquid-vapor interface of a one component fluid. The locations of the surface of tension (SoT) and the Gibbs dividing surface (GDS) are also shown.}
\label{fig:p_dgt}
\end{figure}

Motivated by Eq.~\eqref{eq:p_dgt}, which can be now rewritten as
\begin{equation}\label{eq:p_dgt2}
p_{\scrm{N}}^{\scrm{int}}(z)-p_{\scrm{T}}^{\scrm{int}}(z) = \kappa \left(\dv{\rho}{z}\right)^2, 
\end{equation} 
the pressure anisotropy and the square of the density gradient obtained by MD simulation are compared in Fig.~\ref{fig:p_dgt}. The pressure anisotropy is dominant only in a thin region of several $\sigma$ thickness\cite{Sega2016}.
To the leading order, by counting the number of interactions within the cutoff distance $r_c$, the intermolecular parts of the normal and tangential pressures can be written as
\begin{equation}
\begin{aligned}
p_{\scrm{N}}^{\scrm{int}}(z)
&\propto
-\int_0^{r_c}\dd{z'}\rho(z+z')
\int_0^{r_c-z'}\dd{z''}\rho(z-r'')\\
&\simeq
-\int_0^{r_c}\dd{z'}\int_0^{r_c-z'}\dd{z''}
\left[
\left(\rho(z)+\dv{\rho}{z}z'\right)
\left(\rho(z)-\dv{\rho}{z}z''\right)
\right]\\
&=
-\int_0^{r_c}\dd{z'}\int_0^{r_c-z'}\dd{z''}
\left[
\rho(z)^2-\left(\dv{\rho}{z}\right)^2z'z''
\right]
\end{aligned}
\label{eq:dgt_n}
\end{equation}
and
\begin{equation}
p_{\scrm{T}}^{\scrm{int}}(z)
\propto
-\int_0^{r_c}\dd{x'}\int_0^{r_c-x'}\dd{x''}
\rho(z)^2,
\label{eq:dgt_t}
\end{equation}
considering the symmetry of the system, and therefore the pressure anisotropy can be approximated by
\begin{equation}\label{eq:p_dgt2_model}
p_{\scrm{N}}^{\scrm{int}}(z)-p_{\scrm{T}}^{\scrm{int}}(z)
\propto
\frac{r_c^2}{2}\left(\dv{\rho}{z}\right)^2,
\end{equation}
which reproduces Eq.~\eqref{eq:p_dgt2}. Equation \eqref{eq:p_dgt2_model} tells that the local thermodynamic description of inhomogeneous systems requires the density distribution around the local point, and the point thermodynamic description should fail. 
Considering $(r_c/\sigma)^2/2=6.1$ in the present study, the two profiles in Fig.~\ref{fig:p_dgt} shows a reasonable agreement in the magnitude. The figure shows, however, that there is a clear gap between the peak locations of the two profiles. While the peak location of the density gradient is the location of the GDS [Eq.~\eqref{eq:gds} with $\alpha=1$], the peak location of the pressure anisotropy nearly conincides the location of the SoT given by Eq.~\eqref{eq:sot}. The difference in two peak locations seen in Fig.~\ref{fig:p_dgt} is therefore attributed to the Tolman length, which was $(0.68\pm0.14)\sigma$ for the present case. In deriving Eq.~\eqref{eq:p_dgt2_model}, it was assumed that each pair of molecules within the cut-off range equally contributes to the pressure but it is of course not true. Because the interaction pairs in the liquid phase contribute more to the pressure, the SoT is shifted to the liquid side in comparison to the GDS, which explains the major source of the Tolman length. 

\subsubsection{Two component system}
For the two component fluids, the local pressure tensor $p_{kl}$ is given by
\begin{align}
p_{kl} &= 
\sum_{\alpha=1}^2 
p_{kl,\alpha}^{\scrm{kin}}
+
\sum_{(\alpha,\beta)=(1,1),(1,2),(2,2)} 
p_{kl,\alpha\beta}^{\scrm{int}}\\
p_{kl,\alpha}^{\scrm{kin}} &= 
\frac{1}{S_k \Delta t}
\left<
\sum_{i \in \alpha}^{\Delta t, \scrm{crossing }S_k} \frac{m^i v^i_k v^i_l}{|v^i_k|}
\right>\\
p_{kl,\alpha\beta}^{\scrm{int}} &=
\frac{1}{S_k}
\left<
\sum_{i \in \alpha, j \in \beta}^{\scrm{across }S_k} F^{ij}_l \frac{r^{ij}_k}{|r^{ij}_k|}
\right>,
\end{align}
where $\alpha$ and $\beta$ are fluid component numbers (1 or 2) and the double-counting was avoided for $\alpha=\beta$. For all the components, the kinetic part is isotropic and does not contribute to the pressure anisotropy, as in the one component system. Remembering $p_{\scrm{N}}=p_{zz}$ and $p_{\scrm{T}}=p_{xx}=p_{yy}$, the pressure anisotropy contributed by each component group $(\alpha,\beta)$ is written as
\begin{equation}
p_{\scrm{N},\alpha\beta}(z)-p_{\scrm{T},\alpha\beta}(z)=p_{\scrm{N},\alpha\beta}^{\scrm{int}}(z)-p_{\scrm{T},\alpha\beta}^{\scrm{int}}(z).
\end{equation}
The DGT form of the pressure anisotropy for the two component fluid is obtained by decomposing $\rho=\rho_1+\rho_2$ in Eqs.~\eqref{eq:dgt_n} and \eqref{eq:dgt_t}:
\begin{equation}
p_{\scrm{N},\alpha\beta}^{\scrm{int}}(z)-p_{\scrm{T},\alpha\beta}^{\scrm{int}}(z)
\propto
\omega\frac{r_c^2}{2}
\dv{\rho_{\alpha}}{z}\dv{\rho_{\beta}}{z}
\label{eq:dgt_two}
\end{equation}
for each component group $(\alpha,\beta)$, where $\omega$ is 1 for $\alpha=\beta$ and 2 for $\alpha \neq \beta$. This is the pressure anisotropy expression corresponding to the surface tension formula for multi component systems, shown in Eq.~\eqref{eq:waals_multi}. 

\begin{figure}
\begin{overpic}[width=.95\linewidth]{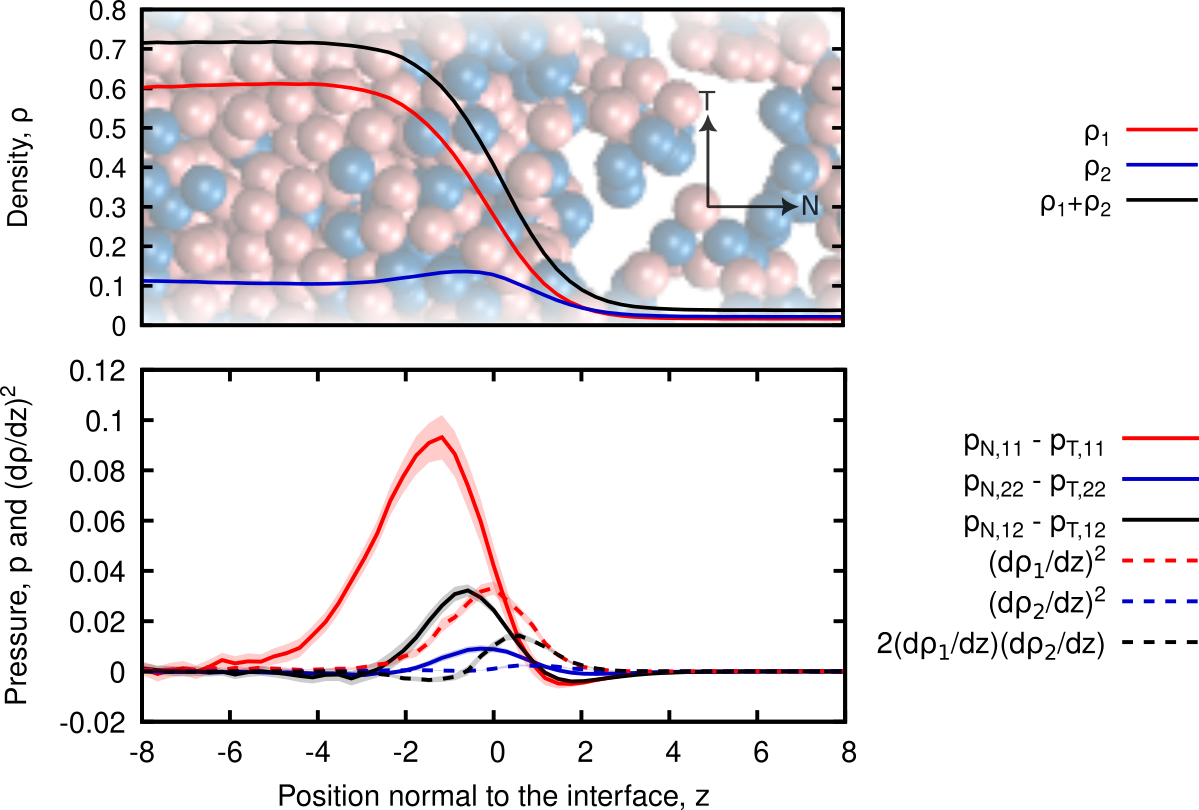}
\put(0,65){(a2)}
\end{overpic}
\begin{minipage}{\linewidth}
\vphantom{1zw}
\end{minipage}
\begin{overpic}[width=.95\linewidth]{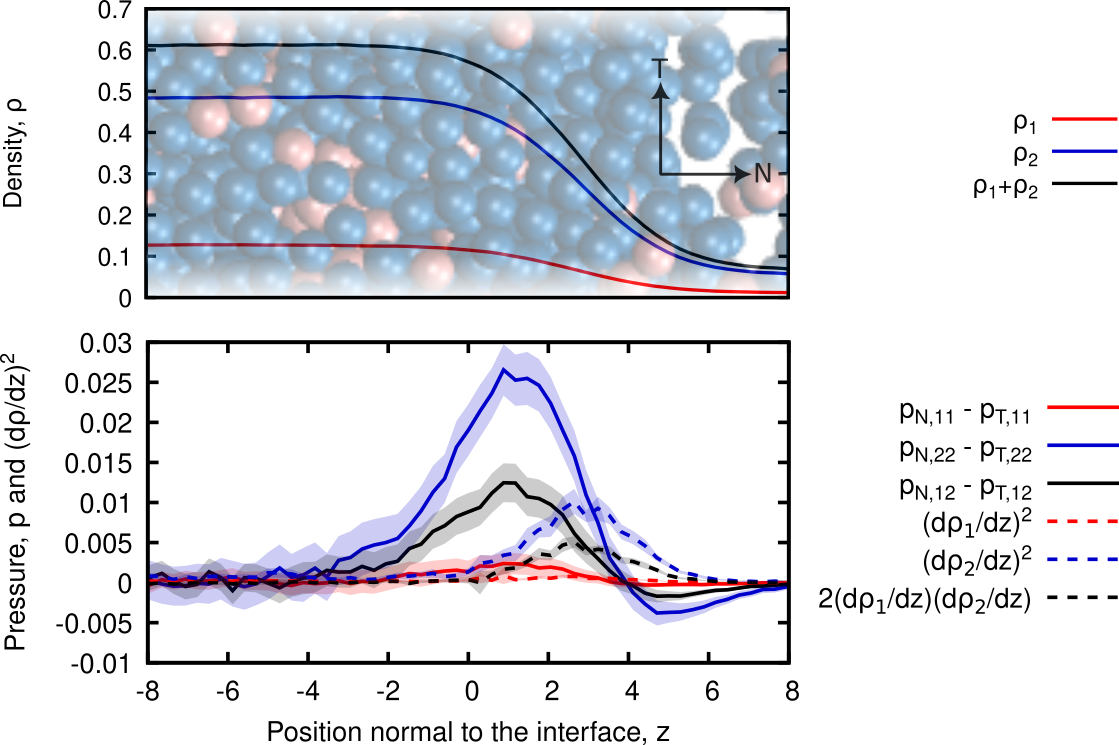}
\put(0,65){(a3)}
\end{overpic}
\caption{Pressure anisotropy and density gradient across the liquid-vapor interface of two component fluids. The total molar fraction of component ``2'' was (a2) 0.2 and (a3) 0.8. (a2) and (a3) correspond to the labels in Fig.~\ref{fig:system}}
\label{fig:p_dgt_two}
\end{figure}
Figure \ref{fig:p_dgt_two} shows the comparison between the pressure anisotropy from each component group and the corresponding density gradients in the liquid-vapor interfaces computed by MD (See a2 and a3 in Fig.~\ref{fig:system}). The total molar fraction of component ``2'', the component of the higher saturation pressure, was 0.2 in Fig.~\ref{fig:p_dgt_two}~(a2) and 0.8 in Fig.~\ref{fig:p_dgt_two}~(a3). 
In both cases, for the interactions in the same components, the relationship between the pressure anisotropy and the density gradient shows the exact similarity to the one component case: it is a reasonable behavior for pair potentials as employed in the present study. For all the component groups the SoT is shifted into the liquid phase due to the same mechanism for the one component case, and the component-wise Tolman length is observed. The small interface adsorption of component ``2" seen in Fig.~\ref{fig:p_dgt_two}~(a2) was also observed by \citet{Stephan2019} for similar LJ parameters. The influence of this adsorption on the pressure anisotropy was marginal.

\subsection{Liquid-liquid interface}
\begin{figure}
\begin{overpic}[width=.95\linewidth]{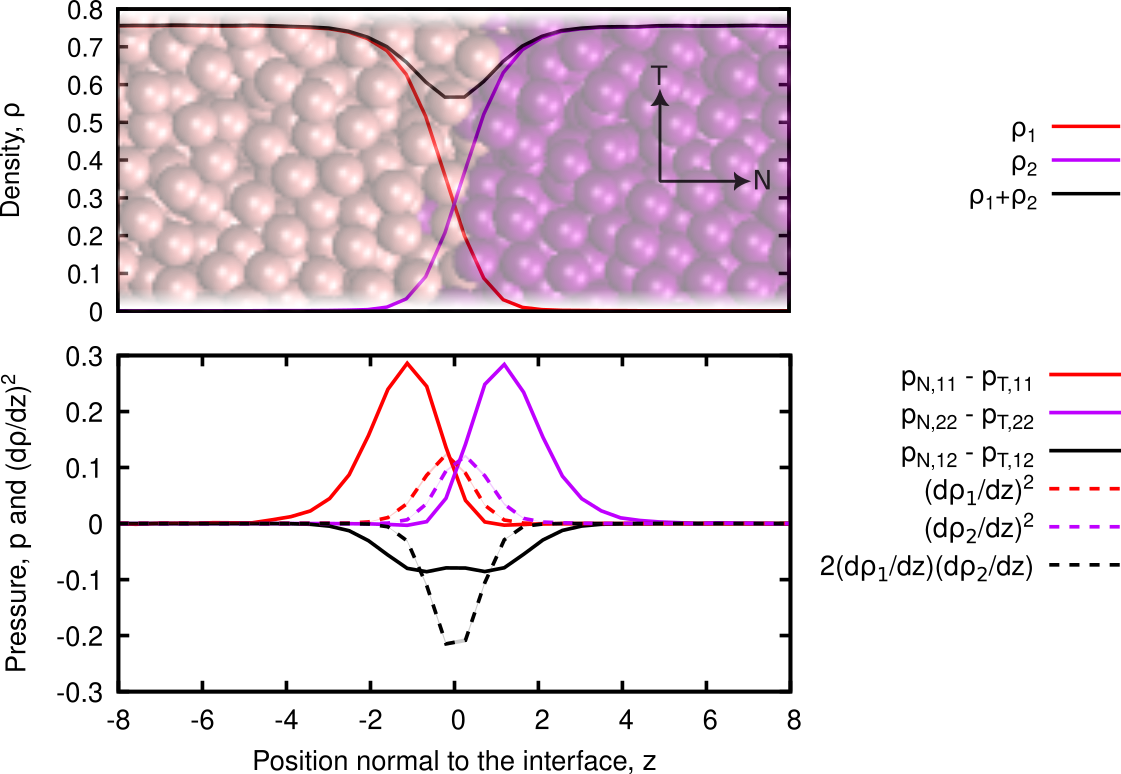}
\put(0,68){(b1)}
\end{overpic}
\begin{minipage}{\linewidth}
\vphantom{1zw}
\end{minipage}
\begin{overpic}[width=.95\linewidth]{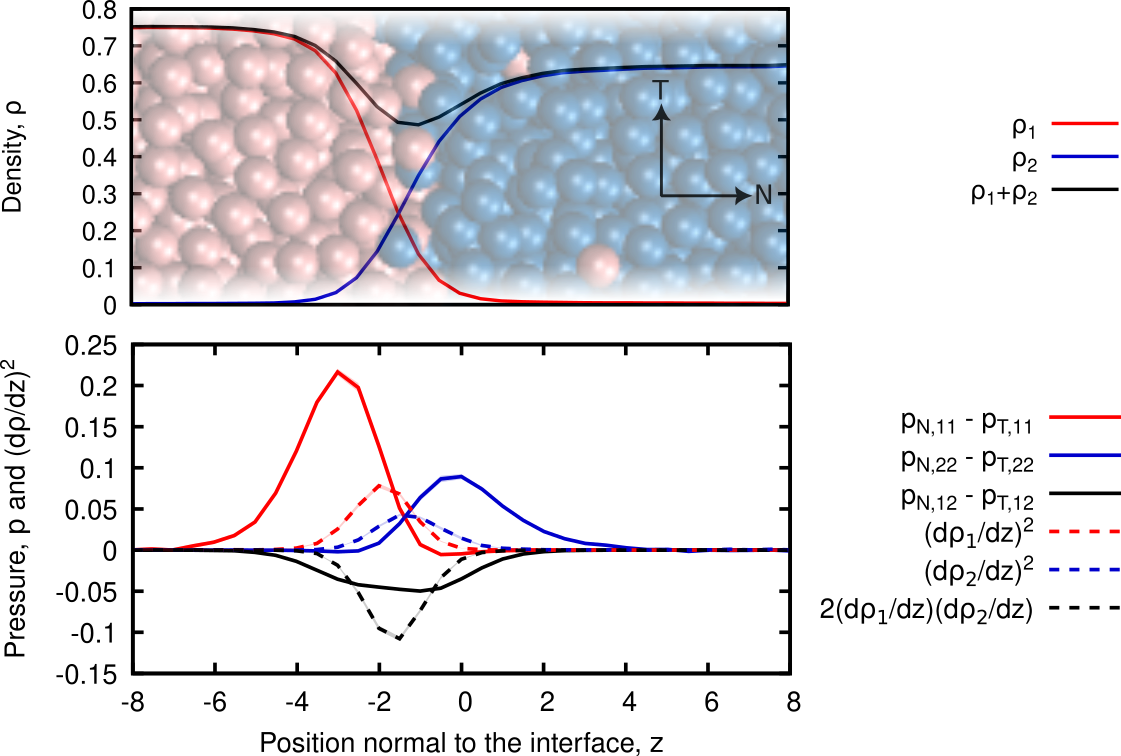}
\put(0,65){(b2)}
\end{overpic}
\caption{Pressure anisotropy and density gradient across the liquid-liquid interface. (b1) and (b2) correspond to the labels in Fig.~\ref{fig:system}}
\label{fig:p_dgt_ll}
\end{figure}

Also for the liquid-liquid interface (See b1 and b2 of Fig.~\ref{fig:system}), the component-wise pressure anisotropy and the density gradients in the interface are compared in Fig.~\ref{fig:p_dgt_ll}. As the top panel of each figure shows, the fluid in the liquid-liquid interface consists of two components, and it is expected that Eq.~\eqref{eq:dgt_two} can be applied here as well. 
The bottom panels show the component-wise breakdown of the pressure anisotropy and the corresponding density gradients in the interface, as shown for the two component liquid-vapor interfaces. As demonstrated in Appendix \ref{app:miscible}, it is not important in terms of local thermodynamics if there is a minimum in the total density distribution between the two liquid slabs. 

The relationship between the pressure anisotropy and the density gradient is similar to the liquid-vapor cases, and the component-wise Tolman length is observed. A distinctive feature in the liquid-liquid interface is that the shift of the SoT relative to the GDS is in the opposite direction for the two components and the (total) Tolman length is much smaller than the liquid-vapor interface.
Another difference from the liquid-vapor interface is that the pressure anisotropy due to the interaction between different fluid components, $p_{\scrm{N},12}-p_{\scrm{T},12}$, gives negative contribution to the surface tension, and it is well described by the cross multiplication of the density gradients, 
\begin{equation}
2\dv{\rho_1}{z}\dv{\rho_2}{z},
\end{equation}
as expected from Eq.~\eqref{eq:dgt_two}. 

\section{Conclusion}
We have reproduced the DGT of van der Waals, as the description of the free energy density of inhomogeneous systems, by assuming that the intermolecular force between the fluid particles in the cut-off range is constant regardless of the interacting distance. By MD simulations, we have shown that the main consequence of this assumption results in the ignorance of the Tolman length, which is the gap between the surface of tension and the Gibbs dividing surface. For multiple component fluids, the Tolman length in the interface is obtained by the average of the component-wise Tolman lengths, which are obtained by the component-wise pressure anisotropy and the corresponding density gradients. 

For inhomogeneous systems, the pressure is a tensor quantity, and the local thermodynamic description should include the pressure tangential to the interface as in Eq.~(\ref{eq:f}) instead of the scalar pressure in homogeneous systems. 
In the free energy density model by the DGT, the tensorial nature of the pressure field is recovered by detecting the interface and its direction in the system by evaluating the density gradients. 
If the Tolman length, which is about the van der Waals radius, is negligibly small compared to the length scale of interest as in the continuum mechanics description of moving interfaces\cite{Jacqmin1996,Qian2006a,Omori2017}, the DGT offers a good basis for the free energy density models for complex inhomogeneous systems, e.g. the Cahn-Hilliard model\cite{cahn1958}. 

\begin{acknowledgments}
All simulations were conducted on SQUID at the Cybermedia Center, Osaka University. The trajectories of the simulated molecular motions were visualized by VMD (Visual Molecular Dynamics).
This work was financially supported by JSPS KAKENHI Grant Nos.\ JP18K03929, JP18K03978, JP22H01400, and JP23H01346. YY was also supported by JST CREST under
Grant No. JPMJCR18I1, Japan.
\end{acknowledgments}

\section*{Author Declarations}
\subsection*{Conflict of Interest}
The authors have no conflicts to disclose.
\section*{Data Availability}
The data that support the findings of this study are available from the corresponding author upon reasonable request.

\appendix
\section{Effect of the confining walls}\label{app:size}
\begin{figure}
\includegraphics[width=.95\linewidth]{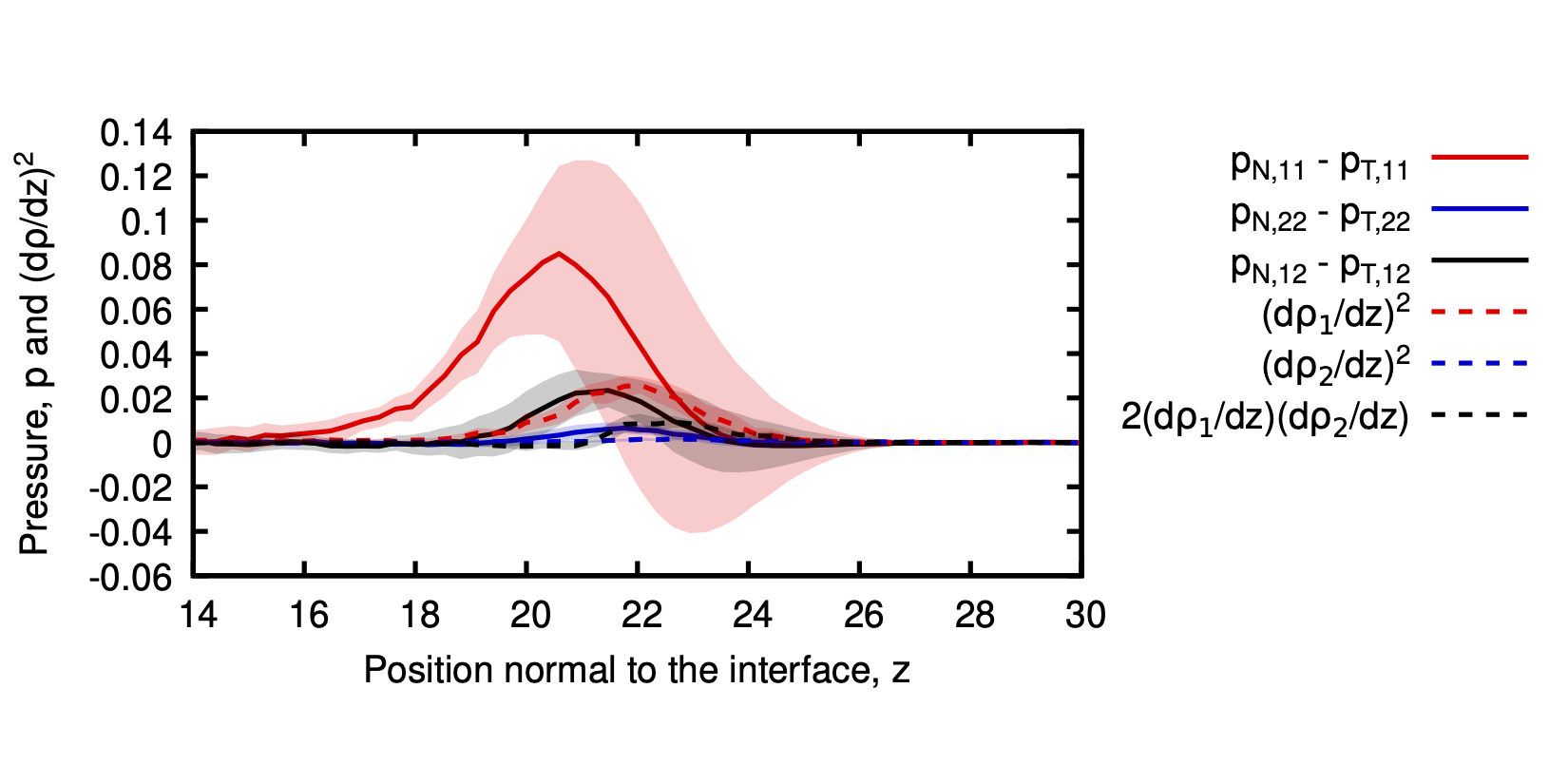}
\caption{Pressure anisotropy and density gradient across the liquid-vapor interface of a two component fluid without confining walls. The LJ potential parameters between fluid particles were identical to (a2) in Fig.~\ref{fig:system}.}
\label{fig:nw}
\end{figure}

To see the effect of the confining walls on the present results, especially on the pressure anisotropy in the interface, two liquid-vapor systems (a1 and a2 in Fig.~\ref{fig:system}) were calculated under periodic boundary condition on all system boundaries. The origin of $z$-axis was adjusted to follow the center of mass of the system. The number of fluid particles and the system size in the $z$ direction were doubled compared to Fig.~\ref{fig:system}, and the statistics were obtained by averaging the results of 20 statistically independent runs.
Figure \ref{fig:nw} shows the pressure anisotropies and the square of the density gradients across the liquid-vapor interface of the two-component fluid: for the one-component fluid, the results were indiscernible from the results with the confining walls. The difference between the results shown in Fig.~\ref{fig:p_dgt_two} (a2) and Fig.~\ref{fig:nw} is no larger than the statistical error. The confining walls suppress the fluctuations in the results but do not bias the statistically averaged results.

\section{Under temperatures closer to the critical point}\label{app:temp}
\begin{figure}
\begin{overpic}[width=.93\linewidth]{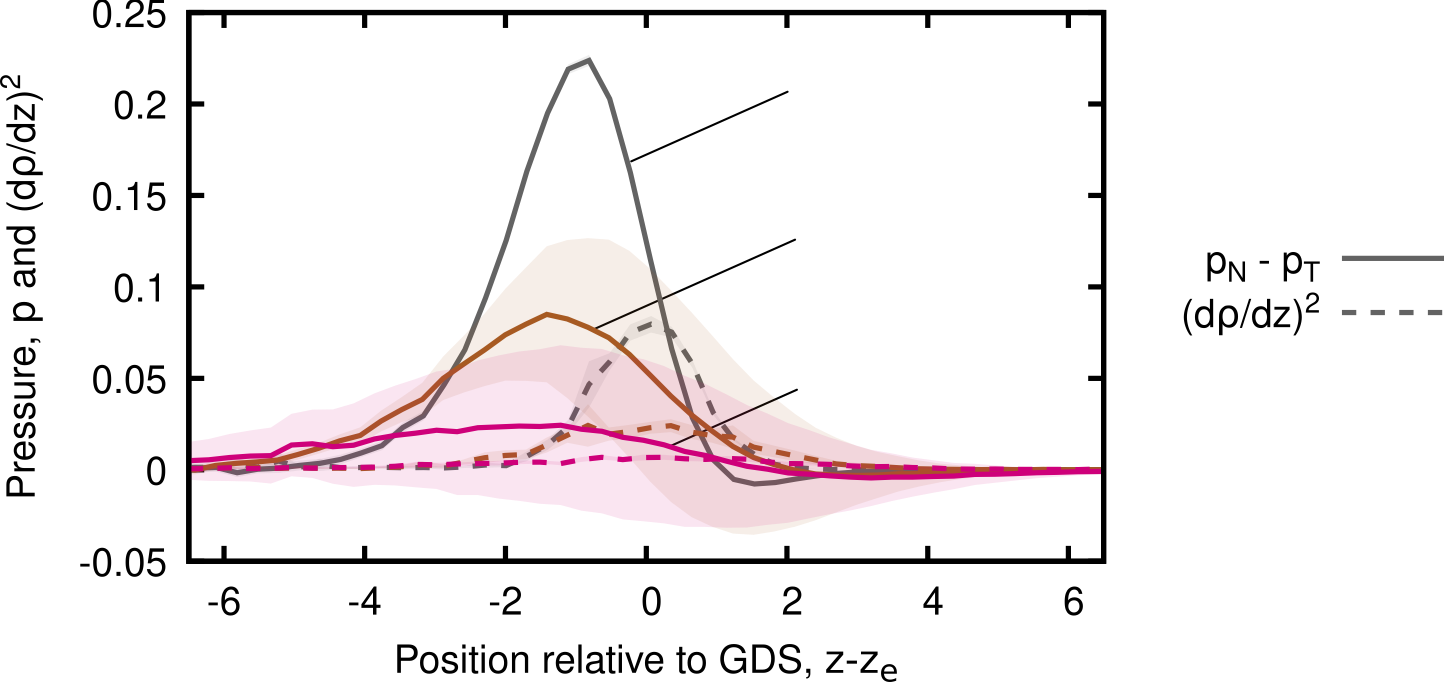}
\put(56,40){$0.83\varepsilon/k_B$}
\put(56,30){$0.95\varepsilon/k_B$}
\put(56,19.5){$1.07\varepsilon/k_B$}
\end{overpic}
\caption{Pressure anisotropy and density gradient across the liquid-vapor interface of a one-component fluid under three temperatures (shown with three different colors).}
\label{fig:diff_temp}
\end{figure}

The DGT is believed to be a better description of inhomogeneous systems under temperatures near the critical temperature, where the density gradient in the interface is small. Figure \ref{fig:diff_temp} shows the pressure anisotropy and the square of the density gradient across the liquid-vapor interface of a one-component fluid under two temperatures $0.95\varepsilon/k_B$ and $1.07\varepsilon/k_B$ in addition to $0.83\varepsilon/k_B$ shown in the main text. The critical temperature of the one-component LJ fluid is $1.31\varepsilon/k_B$\cite{johnson1993,Agrawal1995}. For all three temperatures, Eq.~\eqref{eq:p_dgt2_model} gives a reasonable estimate except the mismatch in the peak locations, which corresponds to the Tolman length. The Tolman lengths $\delta$ were $(0.68\pm0.14)\sigma$, $(0.55\pm0.30)\sigma$ and $(1.43\pm0.50)\sigma$ for temperatures $0.83\varepsilon/k_B$, $0.95\varepsilon/k_B$ and $1.07\varepsilon/k_B$ respectively: $\delta$ tends to increase when the temperature gets closer to the critical temperature\cite{vangiessen2002}. The measured value of $\delta$ may depend on the pressure calculation method. According to Rowlinson and Widom \cite{Rowlinson1982}, $\delta$ measured by the Harasima contour is smaller by about a factor of $3/4$ compared to the IK contour. The difference between the IK contour integral and the MoP is merely in the prefactor in the force summation \cite{shi2023} that reflects the orientation of the connecting vector between the interacting particle locations, and therfore should be smaller than the difference between the two contour methods. 

\section{Pressure anisotropy in the miscible liquid-liquid interface}\label{app:miscible}
\begin{figure}
\includegraphics[width=.95\linewidth]{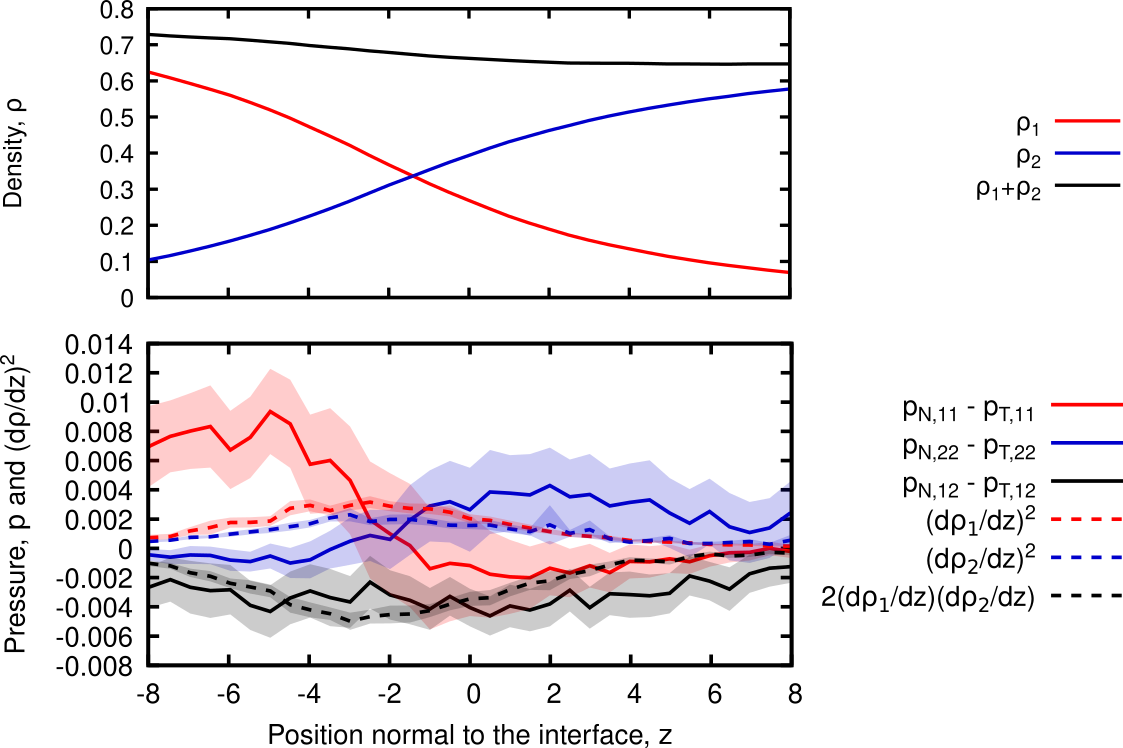}
\caption{Pressure anisotropy and density gradient across the miscible liquid-liquid interface.}
\label{fig:appendix}
\end{figure}

Figure \ref{fig:appendix} shows the density and pressure distributions in the interface between two mutually miscible liquid slabs. For the LJ potentials, the same parameters as the liquid-vapor systems (Tab.~\ref{tab:params}) were employed and the temperature was controlled at $0.83\varepsilon/k_B$. The system setup was similar to (b1) and (b2) in Fig.~\ref{fig:system}, and its height was $49.7\sigma$ so that the system bulk pressure was about $9.4\times10^{-2}\varepsilon/\sigma^3$. The liquid-liquid interface spanned the whole system height. As the top panel of Fig.~\ref{fig:appendix} shows, the total density distribution is monotonous in contrast to Fig.~\ref{fig:p_dgt_ll}. The bottom panel shows that the component-wise pressure anisotropies and density gradients are nevertheless well correlated as in Fig.~\ref{fig:p_dgt_ll}, which indicates that the local thermodynamics of inhomogeneous systems is governed by the density distribution of each component (Eq.~\ref{eq:dgt_two}) but not by the total density distribution.

\bibliography{./zotero.bib}

\begin{thebibliography}{32}
\bibitem [{Notea()}]{Notea}%
  \BibitemOpen
  \bibinfo {note} {From Fig.~\ref {fig:one_c} and Eq.~\protect \textup {\hbox
  {\mathsurround \z@ \protect \normalfont (\ignorespaces \ref
  {eq:p_dgt2}\unskip \@@italiccorr )}}, $p_{\protect \mbox {\relax \protect
  \fontsize {8}{9.5pt}\protect \selectfont T}}$ gets negative when $(\dv *{\rho
  }{z})^2 \gg \rho _g k_B T$, where the RHS is nearly equal to the saturated
  vapor pressure with the vapor density $\rho _g$.}\BibitemShut {Stop}%
\end{thebibliography}

\end{document}